\documentclass[prb,aps,twocolumn,amsmath,amssymb,floatfix,
superscriptaddress]{revtex4}

\date{\today}

\usepackage[dvips]{graphics}
\usepackage{color}
\usepackage{soul}
\definecolor{dred}{rgb}{0,0,0.6}

\begin{document}

\title{Controlled charge and spin current rectifications in a spin 
polarized device}

\author{Moumita Patra}

\affiliation{Physics and Applied Mathematics Unit, Indian Statistical
Institute, 203 Barrackpore Trunk Road, Kolkata-700 108, India}

\author{Santanu K. Maiti}

\email{santanu.maiti@isical.ac.in}

\affiliation{Physics and Applied Mathematics Unit, Indian Statistical
Institute, 203 Barrackpore Trunk Road, Kolkata-700 108, India}

\begin{abstract}

Quasicrystals have been the subject of intense research in the discipline
of condensed matter physics due to their non-trivial characteristic features. 
In the present work we put forward a new prescription to
realize both charge and spin current rectifications considering a 
one-dimensional quasicrystal whose site energies and/or nearest-neighbor 
hopping (NNH) integrals are modulated in the form of well known 
Aubry-Andr\'{e} or Harper (AAH) model, a classic example of an aperiodic 
system. Each site of the chain contains a finite magnetic moment which 
is responsible for spin separation, and, in presence of finite bias an 
electric field is generated along the chain which essentially makes the 
asymmetric band structures under forward and reverse biased conditions, 
yielding finite rectification. Rectification is observed in two forms: 
(i) positive and (ii) negative, depending on the sign of currents in two 
bias polarities. These two forms can only be observed in the case of spin 
current rectification, while charge current shows conventional 
rectification operation. Moreover, we discuss how rectification ratio and 
especially its direction can be controlled by AAH phase which is always 
beneficial for efficient designing of a device. Finally, we critically
examine the role of dephasing on rectification operations. Our study gives 
a new platform to analyze current rectification at nano-scale level, and 
can be verified in different quasi-crystals along with quantum Hall systems.

\end{abstract}

\maketitle

\section{Introduction}

Rectification is one of the fundamental operations in electronic circuits 
and recently a significant attention has been paid to design nano-scale
rectifiers as they are expected to be much more efficient than the 
traditional semiconducting rectifiers. The key idea of having rectification 
is that the current should be different under two biased conditions i.e., 
$I(-V) \ne -I(V)$. This can be done in two ways: (i) by introducing spatial 
asymmetry in the bridging conductor which is the key functional material, 
setting identical conductor-to-electrode coupling, and (ii) by incorporating 
unequal conductor-to-electrode couplings in a spatially symmetric 
conductor~\cite{ref1,ref2,ref3,ref4,ref5,ref6,ref7,ref8,ref9}. The first 
option usually yields better rectification than the other, as in that case 
the density of states spectra in two bias polarities are more distinct. 
A better performance may be expected considering both these two options 
together, though the critical roles played by all other physical factors 
are also quite important for final response.

\begin{figure}[ht]
{\centering \resizebox*{6cm}{3.5cm}{\includegraphics{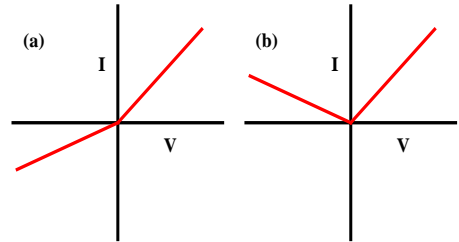}}\par}
\caption{(Color online). Sketches for the (a) positive and (b) negative 
rectifications.}
\label{ws}
\end{figure}

The phenomenon of rectification where we bother only about magnitude of 
currents in positive and negative biases is called as {\em charge current} 
(CC) rectification (CCR). As per the definition of rectification ratio (RR) 
(- current in positive bias/current in negative bias), it is always positive
for CC rectification as we cannot get the currents of same sign under two 
biases. Analogous to CC rectification, there is another type of rectification 
where both magnitude and direction are concerned is known as {\em spin current} 
(SC) rectification (SCR)~\cite{ref10,ref24}. The possible exploitation of
spin degree of freedom triggers us to investigate this SCR phenomenon. This 
is a very new and ongoing field and came into limelight within a decade. 
Unlike CC rectification, SC rectification is rather complex to understand 
since in this case both spin orientations and magnitudes of spin dependent 
currents are involved. Therefore, two kinds of SC rectifications are defined:

\vskip 0.1cm
\noindent
(i) \underline{Positive SC rectification}: Here spin current reverses its sign
under bias inversion, and here we will get positive RR. A sketch of this 
mechanism is given in Fig.~\ref{ws}(a).

\vskip 0.15cm
\noindent
(ii) \underline{Negative SC rectification}: No sign reversal takes place when
bias direction gets altered, as shown in Fig.~\ref{ws}(b), and in this 
situation we will get negative RR.

\vskip 0.15cm
From these two definitions we can see that the positive SC rectification is 
quite similar to CC rectification as sign reversal always takes place under 
bias alteration for the latter one. Now, to achieve SC rectification, the 
symmetry between the spin current components 
needs to be broken, and it is done by considering spin dependent scattering 
mechanisms, along with either of the above two requirements as considered for
CC rectification. For our system, the finite magnetic moments associated with
different lattice sites are responsible for spin separation. As SC 
rectification is closely related to CC rectification, it would be very 
interesting if one can achieve both these two types of rectifications (viz, 
CC and SC) simultaneously, and then a single system can be utilized for 
dual purposes.

Several propositions have been made so far on CC rectification, but too
limited works are available on SC rectification which thus certainly demands
critical analysis to probe into it further. {\em For 
purposeful designing of a 
rectifier we need to focus not only on how to achieve higher rectification
ratio}~\cite{ref1,ref2,ref3,ref4,ref5,ref6,ref7,ref8,ref9,skm1,kwo,kos,skm2}, 
{\em but at the same time emphasis should be given on how its magnitude and 
direction can be controlled selectively}~\cite{ref29,ref29a,ref29b,ref29c}. 
{\em The present work essentially focuses on all these issues.}

The works available in literature are mostly confined within molecular 
systems, quantum dots, grapheme systems, etc.~\cite{ref13,ref14,ref15,ref16,
ref17,ref18,ref19,ref20}, and interest in these systems is gradually dying 
out because of the fact that their performances towards rectification have 
already been revealed, and researchers are trying to find new functional 
elements for fruitful operations.
The recent experimental verification of the existence of non-trivial
topological features~\cite{ref30,ref31} in diagonal and off-diagonal Harper 
models and the equivalence~\cite{topeqi} of these models with Fibonacci and 
other Fibonacci-like quasicrystals have motivated us to test whether any 
non-trivial features are obtained in the rectification
operations or not. Moreover, as we can introduce quasiperiodic modulations
in site energies (diagonal), in nearest-neighbor hopping (NNH) integrals
(off-diagonal), or in both (generalized), we have plenty options to examine
the rectification performance along with transport properties. In order to 
reveal these facts, in the present work, we consider a one-dimensional (1D) 
tight-binding (TB) chain where site energies and/or NNH integrals are 
modulated in the form of Harper model (also called as
Aubry-Andr\'{e} model)~\cite{ref30,ref31,topeqi,ref32,ref33,ref34,ref35}. Each 
site of the chain is accompanied by a magnetic moment (see Fig.~\ref{f0}) 
which is responsible for spin separation. As the spectrum of the system is 
gapped, there is a large possibility to get high degree of spin polarization 
at multiple energy zones associated with the separation of spin channels, 
which is directly reflected in the rectification operation. The gapped 
spectrum has strong effect on CC rectification too. The other important 
aspect of our model is that we have finite possibility to tune both the 
CC and SC RRs by regulating the phase related to site energy, or NNH 
integrals, or by changing both the phases of the generalized Harper model. 
If this tuning mechanism works successfully then it will be very important 
in designing suitable devices.

The rest part of the work is arranged as follows. In Sec. II we discuss 
the model and theoretical prescription for the calculations. All the 
results are critically analyzed in Sec. III, and finally, we conclude 
our essential findings in Sec. IV.

\section{Model and the Method}

\subsection{The Model}

\begin{figure}[ht]
{\centering \resizebox*{8cm}{1.25cm}{\includegraphics{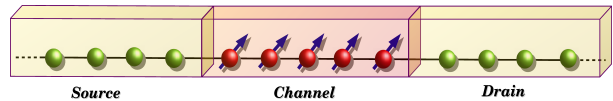}}\par}
\caption{(Color online). Sketch of the nano-junction where a 1D TB AAH chain
(referred as channel) with finite magnetic moments in each lattice sites is 
coupled to two 1D electrodes, source (S) and drain (D). These electrodes are 
semi-infinite, perfect and non-magnetic.}
\label{f0}
\end{figure}

Let us start with the graphical representation of the model nano-junction,
shown in Fig.~\ref{f0}, where a 1D chain is coupled to two semi-infinite 
perfect non-magnetic 1D electrodes, namely, source (S) and drain (D). We 
include AAH modulation in different sectors of the chain, such as site 
energies, or NNH integrals or both, to analyze the precise dependence of 
rectification operation. Each site of this chain is again accompanied by 
a finite magnetic moment as shown by the arrows in Fig.~\ref{f0}.

The Hamiltonian of the nano-junction can be written as a sum of three terms 
\begin{equation}
\mathbf{H} = \mathbf{H_C} + \mathbf{H_{S(D)}} + \mathbf{H_T}
\label{eq1}
\end{equation}
where $\mathbf{H_C}$, $\mathbf{H_{S(D)}}$ and $\mathbf{H_T}$ represent the 
sub-Hamiltonians of the bridging channel, 
source and drain electrodes, and the tunneling coupling between the conducting 
channel and side attached electrodes, respectively. All these Hamiltonians are
described within a TB framework.

The TB Hamiltonian for the chain, with finite magnetic moments at each lattice
sites, considering the modulation both in site energies and NNH integrals reads 
as~\cite{hm1,hm2,hm3},
\begin{eqnarray}
\mathbf{H_C} & = & \sum\limits_{i=1} \mathbf{c_{i,\sigma}
^{\dagger}}\left(\mathbf{\epsilon_{i,\sigma}}-\mathbf{h_i}.
\mathbf{\sigma}\right)\mathbf{c_{i,\sigma}}\nonumber \\
 & & + \sum\limits_{i=1}^{N}\left(\mathbf{c_{i+1,\sigma}^{\dagger}}
\mathbf{t_{i,\sigma}^{\dagger}}\mathbf{c_{i,\sigma}} + 
\mathbf{c_{i,\sigma}^{\dagger}} \mathbf{t_{i,\sigma}}
\mathbf{c_{i+1,\sigma}} \right)
\label{eq2}
\end{eqnarray}
where \\
$\mbox{\boldmath $c_{i,\sigma}$} = \left(\begin{array}{cc}
    c_{i,\uparrow} \\ 
    c_{i,\downarrow}
\end{array}\right)$, $\mbox{\boldmath $c_{i,\sigma}^{\dagger}$}
= \left(\begin{array}{cc}
    c_{i,\uparrow}^{\dagger} & c_{i,\downarrow}^{\dagger}
\end{array}\right)$,
$\mbox{\boldmath $t_{i,\sigma}$}=\left(\begin{array}{cc}
    t_i & 0 \\ 
    0 & t_i
\end{array}\right)$,
$\mbox{\boldmath $\epsilon_{i,\sigma}$}=\left(\begin{array}{cc}
    \epsilon_i & 0 \\ 
    0 & \epsilon_i
\end{array}\right)$,
$\mbox{\boldmath $h_i.\sigma$}=h_i\left(\begin{array}{cc}
    \cos\theta_i & \sin\theta_ie^{-j\varphi_i} \\ 
    \sin\theta_ie^{j\varphi_i} & -\cos\theta_i
\end{array}\right)$.
$c_{i,\sigma}^{\dagger}$ ($c_{i,\sigma}$) is the creation (annihilation) 
operator of an electron at $i$th site with spin 
$\sigma$ ($\uparrow,\downarrow$) and $t_i$ is the NNH integral (i.e., 
hopping between $i$ and $i+1$ ($i-1$) sites). The strength of magnetic 
moment at each site is denoted by $h_i$, and the orientation of any such 
local magnetic moment is described by the polar angle $\theta_i$ and 
azimuthal angle $\varphi_i$ as used in conventional polar coordinate system.

The modulations in site energies and NNH integrals are chosen 
as~\cite{ref33,ref34,ref35}
\[
\epsilon_i=\nu \cos\left(2 \pi b i + \phi_{\nu}\right)~~\mbox{and}~~
t_i=1 + \lambda \cos\left(2 \pi b i + \phi_{\lambda}\right)
\]
where $b$ is an irrational number and we choose it as the golden mean,
$\nu$ and $\lambda$ are the modulation strengths, and $\phi_{\nu}$ and 
$\phi_{\lambda}$ are the AAH phases, which can be tuned independently 
with suitable setup.

Now, as voltage bias is applied between the electrodes, an electric field
is established which in turn modifies the site potentials. Therefore,
we can write the effective site energy as a sum of two 
terms~\cite{ref24,efld1,efld2}
\begin{equation}
\epsilon_i^{eff}=\epsilon_i^{(0)} + \epsilon_i(V)
\label{eq4}
\end{equation}
where $\epsilon_i^{(0)}$ is the voltage independent term, and for the
diagonal Harper model it becomes identical to that what is described above
for $\epsilon_i$. The voltage dependent term $\epsilon_i(V)$ is rather 
very hard to determine from first principle calculations as it involves 
complex many-body solutions. Therefore, in our work, heuristically we can 
consider a potential profile in the form of a linear bias drop. This is
reasonably a good choice and one can get the physical essence of rectification
operation very nicely. One may also choose other potential profiles, but 
the fact is that all the physical pictures will remain same qualitatively.
Thus, as a matter of simplification we consider only the linear bias drop 
along the chain, and we can express the profile for a $N$-site chain 
as~\cite{ref24,efld1,efld2} $\epsilon_i(V)=V/2-iV/(N+1)$, where $V$ is the 
bias drop across the junction.

The other two sub-Hamiltonians of Eq.~\ref{eq1}, $\mathbf{H_{S(D)}}$ and
$\mathbf{H_T}$, will have the very simple TB forms as the electrodes are 
perfect and non-magnetic. These electrodes are parameterized by on-site 
energy $\epsilon_0$ and NNH integral $t_0$, and they are directly coupled 
at the two ends of the magnetic channel with the coupling strengths $t_S$ 
and $t_D$. 

\subsection{The Method}

The common quantity that is required to describe transport properties is the 
transmission function. We evaluate it using wave-guide (WG) theory, a standard 
technique~\cite{wg1,wg2,wg3,wg4} for calculating transmission probability. 
One can also use some other prescriptions like transfer-matrix formalism
or Green's function technique~\cite{tm1,tm2,gn1,gn2,gn3,gn4}. Now, in the WG 
method, a set of coupled
linear equations involving wave amplitudes at different lattice sites of
the chain along with the boundary sites of the electrodes with which the
chain is coupled are solved. Considering plane wave incidence of up and
down spin electrons with unit amplitude from the source electrode, we 
solve the coupled equations to find the spin dependent reflection and 
transmission amplitudes, $r_{\sigma \sigma^{\prime}}$ and 
$\tau_{\sigma \sigma^{\prime}}$, respectively. Using these quantities,
we calculate reflection and transmission probabilities as 
$R_{\sigma \sigma^{\prime}}=|r_{\sigma \sigma^{\prime}}|^2$ and
$T_{\sigma \sigma^{\prime}}=|\tau_{\sigma \sigma^{\prime}}|^2$.
A detailed theoretical prescription of the WG formalism is given in 
Appendix~\ref{transp}. Up to now, effect of dephasing is not included in 
the calculations, and the inclusion of it can be understood from the other 
part of our work. 

Once the transmission function $T_{\sigma \sigma'}$ is obtained, the
spin dependent junction current is computed from the relation~\cite{gn2,gn3}
\begin{equation}
I_{\sigma \sigma'} = \frac{e}{h} \int dE \, T_{\sigma \sigma'}(E) 
\left[f(E-\mu_S)-f(E-\mu_D)\right]
\label{eq5}
\end{equation}
where $f$ is the Fermi-Dirac distribution function, $\mu_S$ and $\mu_D$ 
($= E_F \pm eV/2$) are the electro-chemical potentials of S and D, respectively,
and $E_F$ represents the equilibrium Fermi energy. As thermal broadening is
too weak compared to the broadening caused by the conductor-to-electrode
coupling, we can safely ignore the effect of temperature, and thus, throughout
the analysis we set the system temperature to zero, without loss of any
generality. Under this assumption the above current expression boils down 
to~\cite{gn2,gn3} 
\begin{equation}
I_{\sigma \sigma'}(V) = \frac{e}{h} \int\limits_{E_F-\frac{eV}{2}}^{E_F+
\frac{eV}{2}}T_{\sigma \sigma'}(E) \, dE
\label{eq6}
\end{equation}
From Eq.~\ref{eq6} we calculate all the spin dependent currents at required
bias voltages and then determine the junction charge and spin currents using
the definitions $I_c=I_{\uparrow} + I_{\downarrow}$ and $I_s=I_{\uparrow} -
I_{\downarrow}$, respectively. We refer $I_{\uparrow}=I_{\uparrow\uparrow} 
+ I_{\downarrow\uparrow}$ and $I_{\downarrow}=I_{\downarrow\downarrow}
+ I_{\uparrow\downarrow}$.

Finally, we define the charge and spin current rectification ratios 
as~\cite{ref24}
\begin{eqnarray}
RR_c & = & -\frac{I_c(+V)}{I_c(-V)}~~ \mbox{and}~~ 
RR_s=-\frac{I_s(+V)}{I_s(-V)}
\label{eq7}
\end{eqnarray}
$RR_{c(s)}=1$ means no rectification. As the rectification is measured by the
ratios of currents in two bias polarities, sometimes it is very hard to read,
and therefore we also calculate inverse of it i.e., $1/RR_{c(s)}$ along with 
$RR_{c(s)}$.

\begin{figure*}[ht]
{\centering \resizebox*{8cm}{8.5cm}{\includegraphics{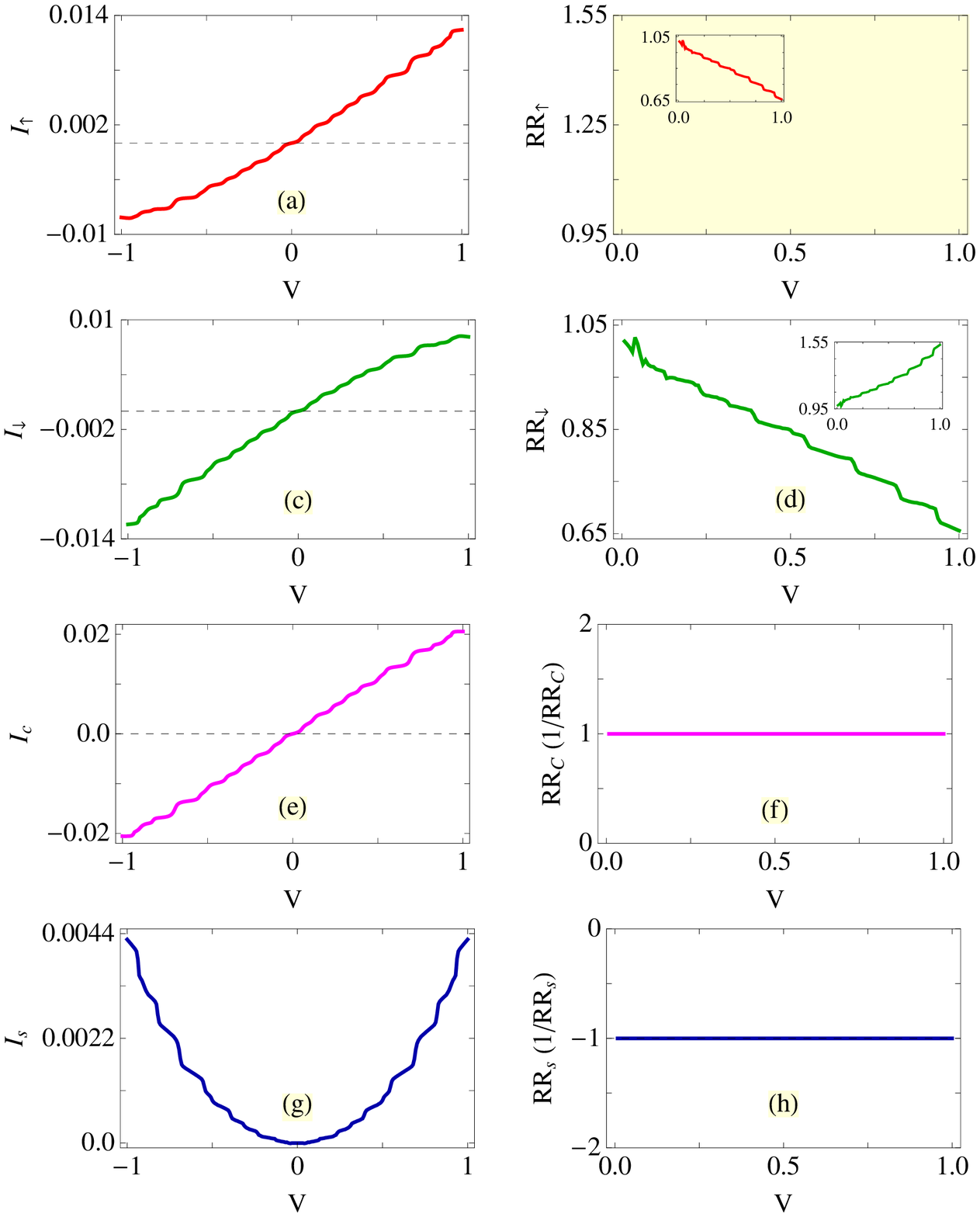}}
\resizebox*{8cm}{8.5cm}{\includegraphics{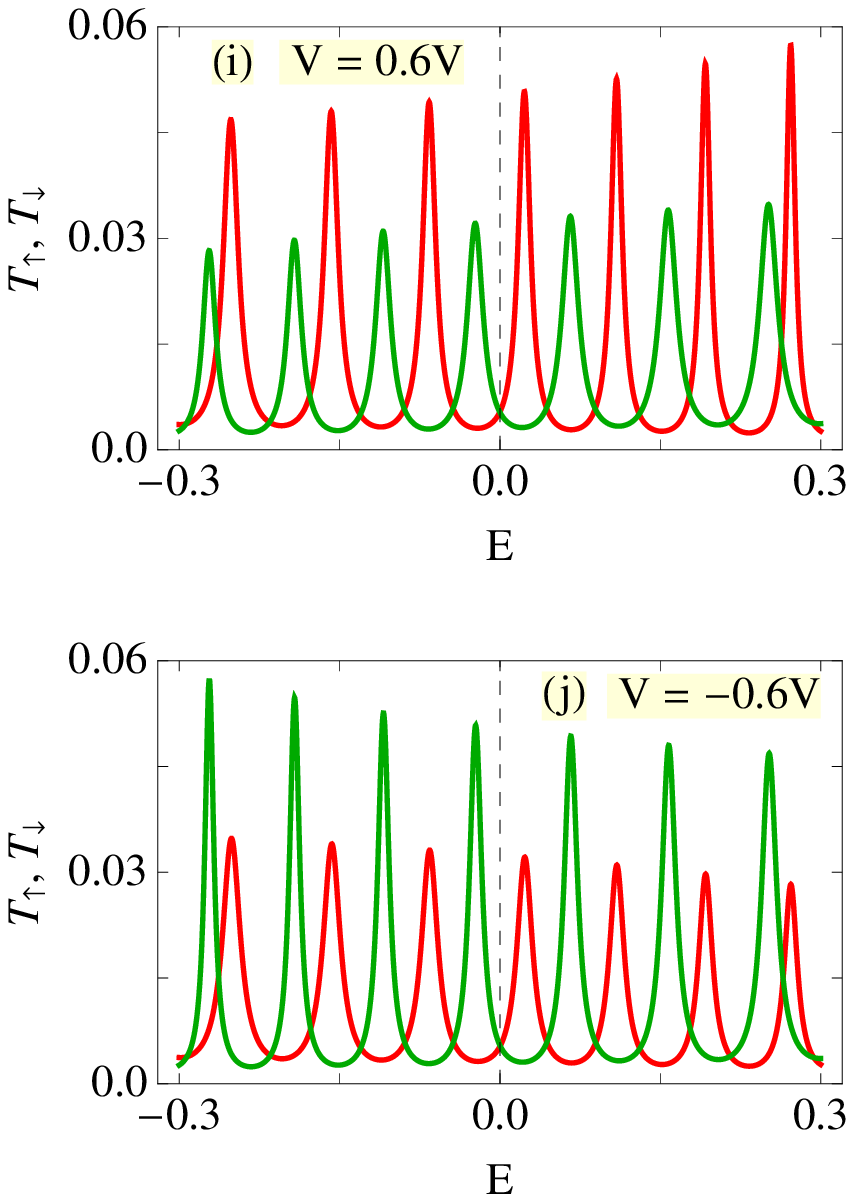}}\par}
\caption{(Color online). Different currents and rectification ratios (a-h) as a 
function of bias voltage for a perfect chain considering $N=60$, $h=1$ and 
$\nu=\lambda=0$. In the insets of (b) and (d), the inverse of RR is shown. In
(i) and (j), spin dependent transmission probabilities as a function of energy
in two bias polarities are shown for a typical bias voltage $V=0.6\,$V, where 
the red and green lines correspond to the transmission probabilities for up and 
down spin electrons, respectively.}
\label{fig3}
\end{figure*}

\begin{figure*}[ht]
{\centering \resizebox*{10cm}{9cm}{\includegraphics{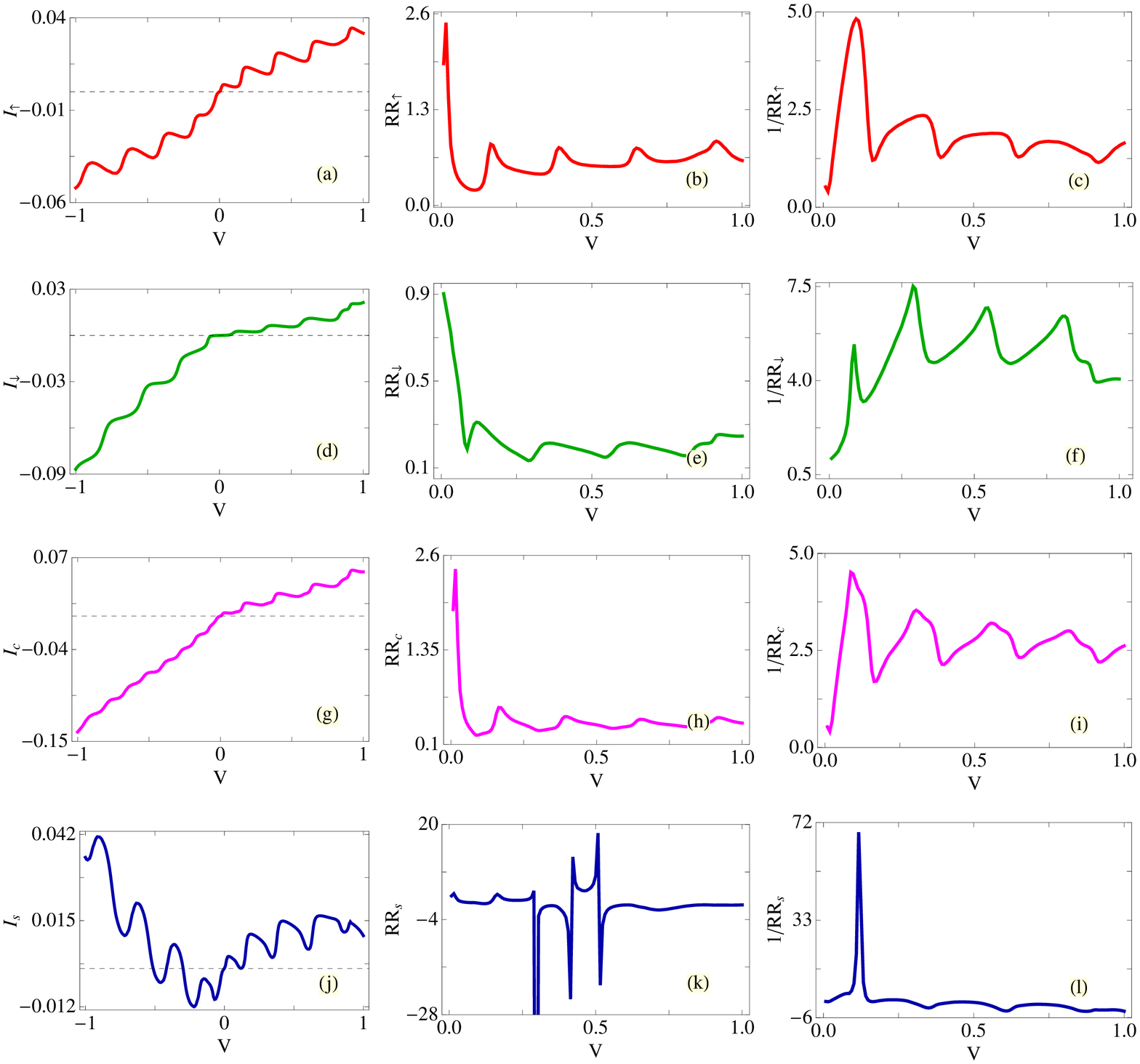}}
\raise 1cm \hbox {\kern 0.1cm \resizebox*{6cm}{7cm}
{\includegraphics{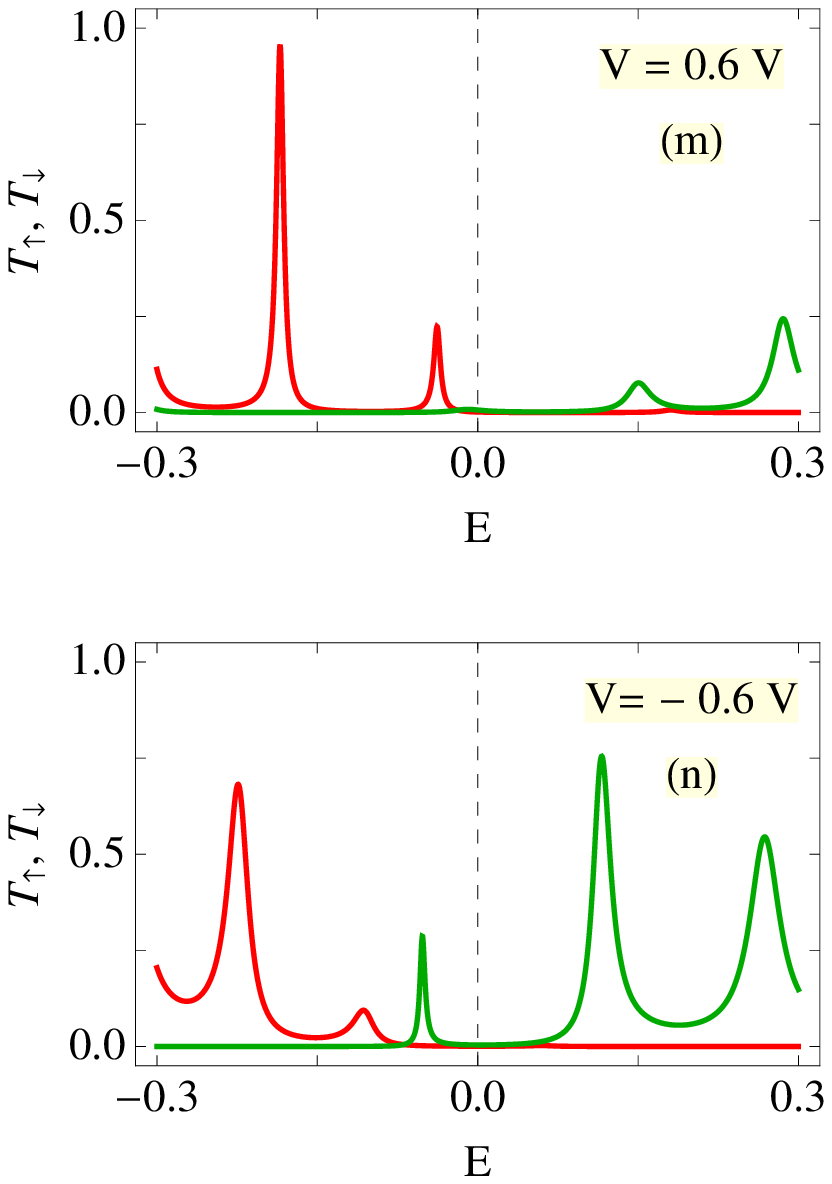}}}\par}
\caption{(Color online). Different currents and rectification ratios as a 
function of bias voltage along with spin dependent transmission probabilities 
for the 1D diagonal AAH chain considering $N=40$, $\nu=0.5$, $\phi_{\nu}=0$, 
$\lambda=0$, and $h=0.5$. The red and green lines in (m) and (n) represent 
the identical meaning as described in (i) and (j) of Fig.~\ref{fig3}.}
\label{fig4}
\end{figure*}

\section{Results and discussion}

Now, we present our essential results. The common parameter values used to 
carry out numerical calculations are as follows. For the side-attached
1D electrodes we choose $\epsilon_0=0$ and $t_0=3$. The magnetic moments 
in the bridging chain are assumed to be aligned along the positive 
$Z$-direction i.e., $\theta_i=\varphi_i=0\,\forall\,i$. The 
conductor-to-electrode coupling strengths are chosen as asymmetric, 
$t_S=0.4$ and $t_D=1$, and the equilibrium Fermi energy $E_F$ is fixed 
to zero. All the energies are measured in unit of electron-volt (eV), and 
currents are computed in unit of ($e/h$). Unless otherwise stated, we do 
not consider the effect of dephasing right now, and we discuss it at the 
end of our analysis in a separate sub-section.

\subsection{Ordered chain}

Before addressing the central results i.e., precise roles of quasiperiodic 
modulations on rectifications, let us first focus on the rectification 
operation 
considering a perfect chain ($\epsilon_i^{(0)}=0\,\forall\,i$) for the sake 
of illustration and to understand the basic mechanisms. The results computed 
for a $60$-site chain are shown in Fig.~\ref{fig3}, where we present spin 
dependent currents along with charge and spin currents, rectification ratios 
and the two-terminal transmission probabilities of up and down spin electrons.
Several important features are observed. The individual spin current components
(up and down) get unequal magnitudes (Figs.~\ref{fig3}(a) and (c)) in 
two bias polarities. Therefore, finite rectification for these two spin 
currents are obtained as shown in Figs.~\ref{fig3}(b) and (d). Now, looking 
carefully into the variations of up and down spin currents with voltage bias 
we see that the nature of up spin current in one bias polarity gets exactly 
reversed in the case
of down spin current under bias reversal. Due to this fact, $RR_{\sigma}$ 
becomes identical with $1/RR_{\sigma^{\prime}}$ (see Figs.~\ref{fig3}(b), (d) 
and their insets). From the characteristics of spin dependent currents and the
corresponding rectifications, we can now easily get the dependence of charge
and spin currents and the associated rectifications. Both for the charge and
spin currents the magnitudes are exactly identical in two biased conditions,
where the usual phase reversal is obtained in charge current, and for the
case of spin current no sign alteration takes place. Accordingly, no 
rectification
is available for charge current ($RR_c=1$), whereas spin current provides 
$RR_s=-1$. We call it (viz, $RR=-1$) as {\em full wave rectification}. Here 
it is 
relevant to note that, one can get rectification, as mentioned earlier, either 
by considering a spatially asymmetric conductor setting identical 
conductor-electrode coupling ($t_S=t_D$), or by considering unequal coupling
($t_S \ne t_D$) for a spatially symmetric conductor or by both. But, we
see that no rectification is available for $I_c$ though we set asymmetric 
couplings of the conductor to the side attached electrodes. The reason is that 
we set the voltage independent site energies ($\epsilon_i^{(0)}$) to zero. 
Instead of this zero, if one takes any finite value then the charge current 
will exhibit finite rectification under this condition.

The above phenomena can be explained as follows. Let us look into the spectra 
given in Figs.~\ref{fig3}(i) and (j) where we plot the transmission probabilities
of up (red line) and down (green line) spin electrons for a typical bias voltage
under its two polarities. Sharp resonant peaks are observed associated with the
resonant energies. The interesting feature is that a perfect swapping of both the 
magnitude and phase takes place between the transmission probabilities of two 
spin components under bias alteration. It happens as we choose a perfect conductor.
Now, the sign and magnitudes of individual currents (up and down), charge and 
spin currents as well as $RR_{c(s)}$ can be easily understood since current 
involves the integration of the transmission function (Eq.~\ref{eq6}). Suppose 
the areas under the spectra $T_{\uparrow}(V)$-$E$ and $T_{\downarrow}(-V)$-$E$ are 
identical (as they should be) to $x$ for any typical bias $V$, and the areas
under the curves $T_{\downarrow}(V)$-$E$ and $T_{\uparrow}(-V)$-$E$ are 
identical to $y$. Then the transport charge current $I_c$ for positive 
bias should be $I_c(V)=I_{\uparrow}(V) + I_{\downarrow}(V)=(e/h)(x+y)$.
Similarly, in the negative bias condition it becomes 
$I_c(-V)=I_{\uparrow}(-V) + I_{\downarrow}(-V)=-(e/h)(x+y)=-I_c(V)$.
Thus, getting of the identical magnitude with sign reversal for charge current
under two bias polarities is clearly understood. In the same fashion we can
see that $I_s(-V)=I_s(V)$ i.e., no phase reversal takes place for the
spin current yielding $RR_s=-1$.

\subsection{Diagonal AAH chain}

Following the above analysis now we can explore the critical roles played by
\begin{figure}[ht]
{\centering \resizebox*{7cm}{8cm}{\includegraphics{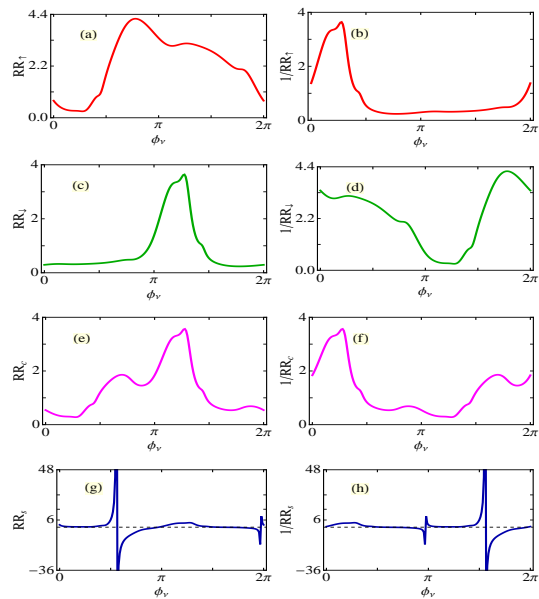}}\par}
\caption{(Color online). Tuning of rectification ratios by the diagonal AAH phase 
$\phi_{\nu}$ for a diagonal AAH chain. Here we choose $N=50$, $\nu=0.5$, $h=0.5$
and $\lambda=0$. The results are computed at $V=0.5\,$V.}
\label{fig5}
\end{figure}
quasiperiodic modulations, in different forms, on rectifications those have not
been addressed earlier in literature. To explore these facts, let us begin by 
considering the modulation in the diagonal part. The results are shown in 
Fig.~\ref{fig4} for a $40$-site chain where we compute different currents, 
rectification ratios together with spin dependent transmission probabilities.
In presence of aperiodic site energies the symmetry between two
spin currents with bias alteration no longer persists, unlike the case what 
we get in the perfect chain. The suitable hint of getting different current 
amplitudes in positive and negative biases can be obtained from the nature of
the transmission spectra (Figs.~\ref{fig4}(m) and (n)), computed at a typical 
bias voltage. The areas under the red curves for both $V=\pm 0.6\,$V are quite
comparable, whereas they differ reasonably well for the green ones. These are
exactly reflected in the spin dependent currents (see Figs.~\ref{fig4}(a) and
(d)). For this diagonal AAH model, the magnitudes of charge current in two 
bias polarities are different (Fig.~\ref{fig4}(g)) which results a finite 
rectification, unlike the ordered chain, as shown in Fig.~\ref{fig4}(h). From
this figure apparently we see that $RR_c$ reaches to $\sim 2.6$ at one particular 
(low) voltage, while it becomes too small for all other voltages. This is due to
the fact that $|I_c(-V)|$ is quite large compared to the $|I_c(V)|$ 
(Fig.~\ref{fig4}(g)), and therefore when we take the inverse of $RR_c$, we 
find moderate values, as presented in Fig.~\ref{fig4}(i). Thus, both $RR$ and
$1/RR$ are required to analyze to have the complete picture of rectification.
The behavior of spin current and its rectification is quite interesting (see
Figs.~\ref{fig4}(j)-(l)). Both the positive and negative spin currents are now
obtained in two bias polarities, unlike the charge current, and therefore, two
types of rectifications (positive and negative) are available. Also the degree 
of rectification is too large that is one of our primary requirements. 

Now, to examine how the rectification operation gets changed with the modulation
of quasiperiodic site energies, in Fig.~\ref{fig5} we show the dependence of $RR$
on the diagonal AAH phase $\phi_{\nu}$, as this phase directly modulates the site 
energies. The results are computed for a typical bias voltage $V=0.5\,$V 
considering a $50$-site diagonal AAH chain. We find a very strong dependence of
$\phi_{\nu}$ on rectification. Both for the up and down spin currents, the
rectification ratio and its direction can be changed widely by varying the phase 
$\phi_{\nu}$.
As a results of this, we get significant variation in charge and spin current
rectifications. Most importantly, as this phase factor can be regulated 
{\em externally} with a suitable setup, we get a suitable hint of designing 
externally controlled efficient rectifier at nanoscale level with quasiperiodic 
systems. The underlying physics is that, the gapped energy spectrum of the 
quasiperiodic system is modified with the phase factor, which thus may produce
more asymmetric density of states spectra under two different bias polarities,
\begin{figure}[ht]
{\centering \resizebox*{8cm}{6cm}{\includegraphics{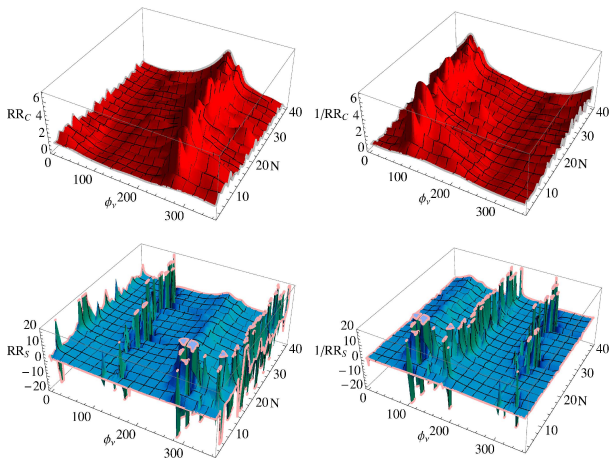}}\par}
\caption{(Color online). Simultaneous variation of rectification ratios with 
system size $N$ and AAH phase $\phi_{\nu}$ for a diagonal Harper chain. Here 
we set $\nu=0.5$, $\lambda=0$, $h=0.5$ and $V=0.5\,$V.}
\label{fig6}
\end{figure}
resulting higher rectification. Along with these characteristics we would like 
to note another important feature that for the situation where $RR$ (or $1/RR$) 
is too large, then the current in any one of the two biased conditions is 
very high than the other one which yields {\em half wave rectification}.

The results studied above are worked out for some typical chain lengths. In
order to characterize the results for other systems sizes and at the same time
to test whether there is any correlation between system size $N$ and the AAH 
phase $\phi_{\nu}$, in Fig.~\ref{fig6} we present the variation charge and
spin current rectifications as functions of $N$ and $\phi_{\nu}$. The results
are somewhat interesting and important too. At a first glance we see that 
charge current provides reasonably good rectification (the maximum of $RR_c$
reaches very close to $6$), while for the other current (spin), the ratio
is too high at some typical phases that we truncate the peaks after a certain
limit for better viewing of the spectra. These high peaks essentially correspond
to the half wave rectification. The other important signature is that, the 
results are mostly affected by the phase factor, rather than the system size.
For a fixed $\phi_{\nu}$, $RR_{c(s)}$ is almost invariant with $N$, while 
change of $\phi_{\nu}$ leads to a dramatic change. Thus, we can argue that
the results are robust and can be checked for a wide range of $N$. 

\subsection{Off-diagonal AAH chain}

Now we consider the system with AAH modulation in NNH integrals, keeping
the diagonal part free from any aperiodicity. In the diagonal part we set 
$\epsilon_i^{(0)}=0$. For this configuration, the results are rather less
interesting, analogous to the perfect chain. If we look into the spectra
\begin{figure}[ht]
{\centering \resizebox*{8cm}{3cm}{\includegraphics{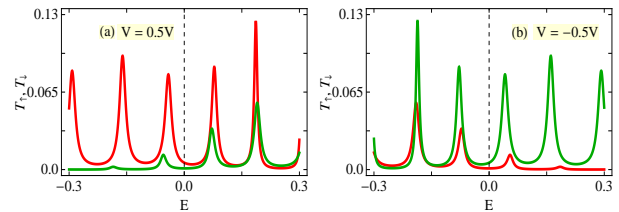}}\par}
\caption{(Color online). $T_{\uparrow}$-$E$ and $T_{\downarrow}$-$E$ 
characteristics for the off-diagonal AAH chain in two bias polarities at a 
typical bias voltage $V=0.5\,$V. The other physical parameters are: $\nu=0$, 
$\lambda=0.5$, $\phi_{\lambda}=0$, $N=40$ and $h=0.5$.}
\label{fig7}
\end{figure}
\begin{figure}[ht]
{\centering \resizebox*{8cm}{6cm}{\includegraphics{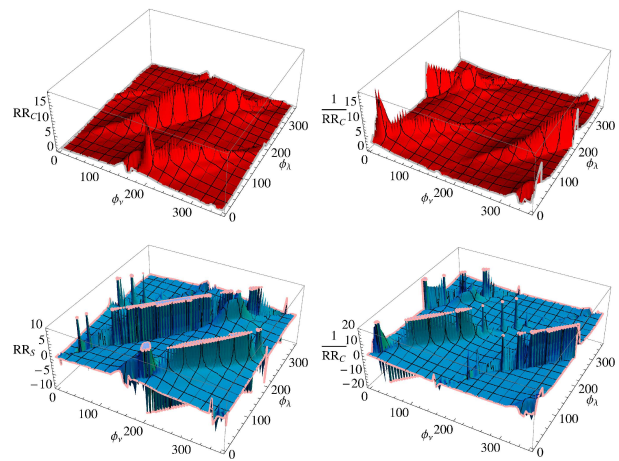}}\par}
\caption{(Color online). Simultaneous variation of rectification ratios with 
diagonal and off-diagonal AAH phases, $\phi_{\nu}$ and $\phi_{\lambda}$, for a 
generalized AAH chain. Here we set $\nu=0.5$ $\lambda=0.25$, $h=0.5$, $N=40$
and $V=0.3\,$V.}
\label{fig8}
\end{figure}
given in Fig.~\ref{fig7}, we can see that the transmission probabilities of 
up and down spin electrons get exchanged under swapping the bias polarities.
As a results of this, we cannot expect any rectification in charge current
($RR_c=1/RR_c=1$), and for spin current the rectification ratio is always 
identical to $-1$ and there is no question about the tuning of $RR_s$ by
means of the off-diagonal AAH phase $\phi_{\lambda}$.

\begin{figure*}[ht]
{\centering \resizebox*{3.5cm}{7cm}{\includegraphics{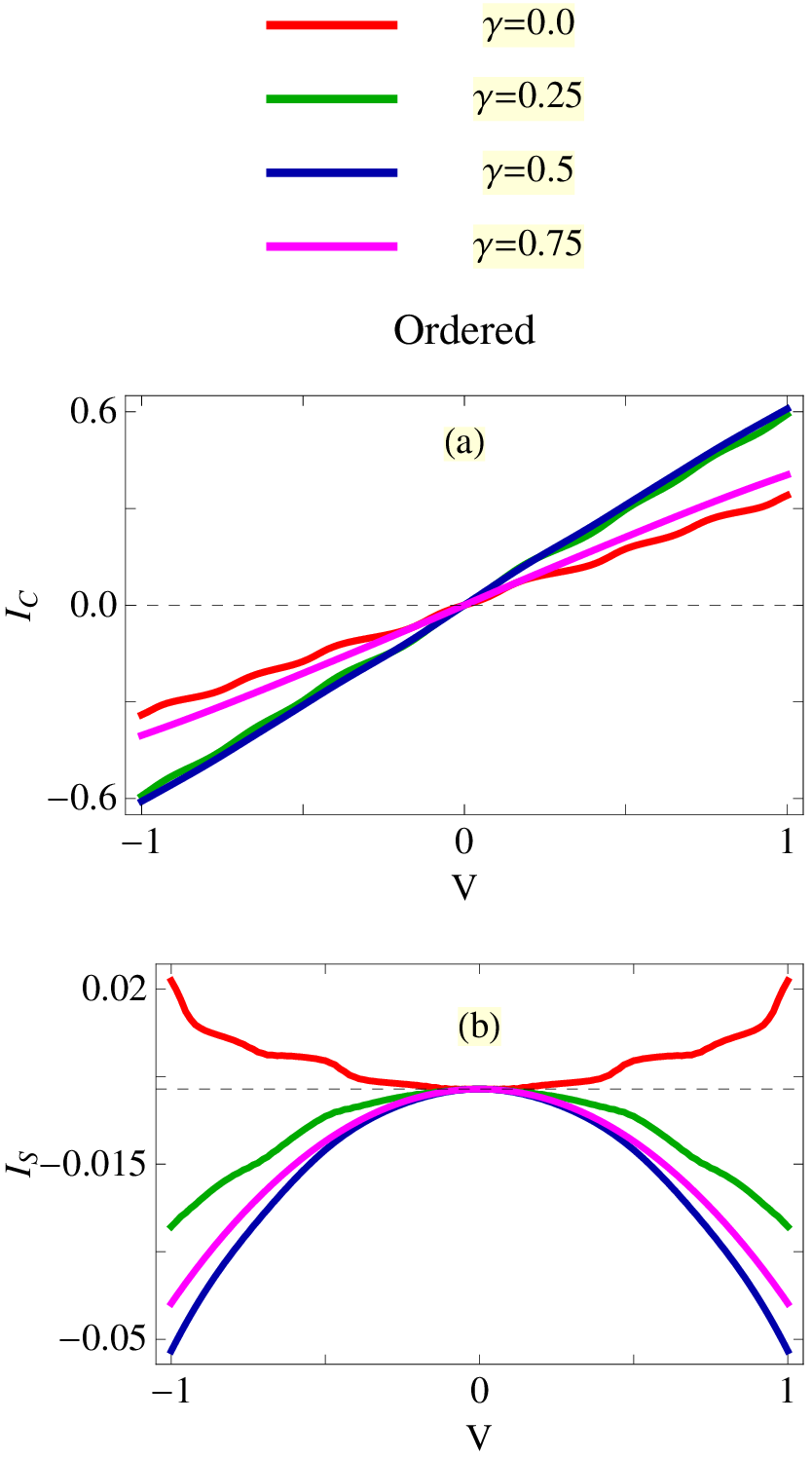}}
\raise 0cm \hbox {\kern 0.5cm \resizebox*{11cm}{7cm}
{\includegraphics{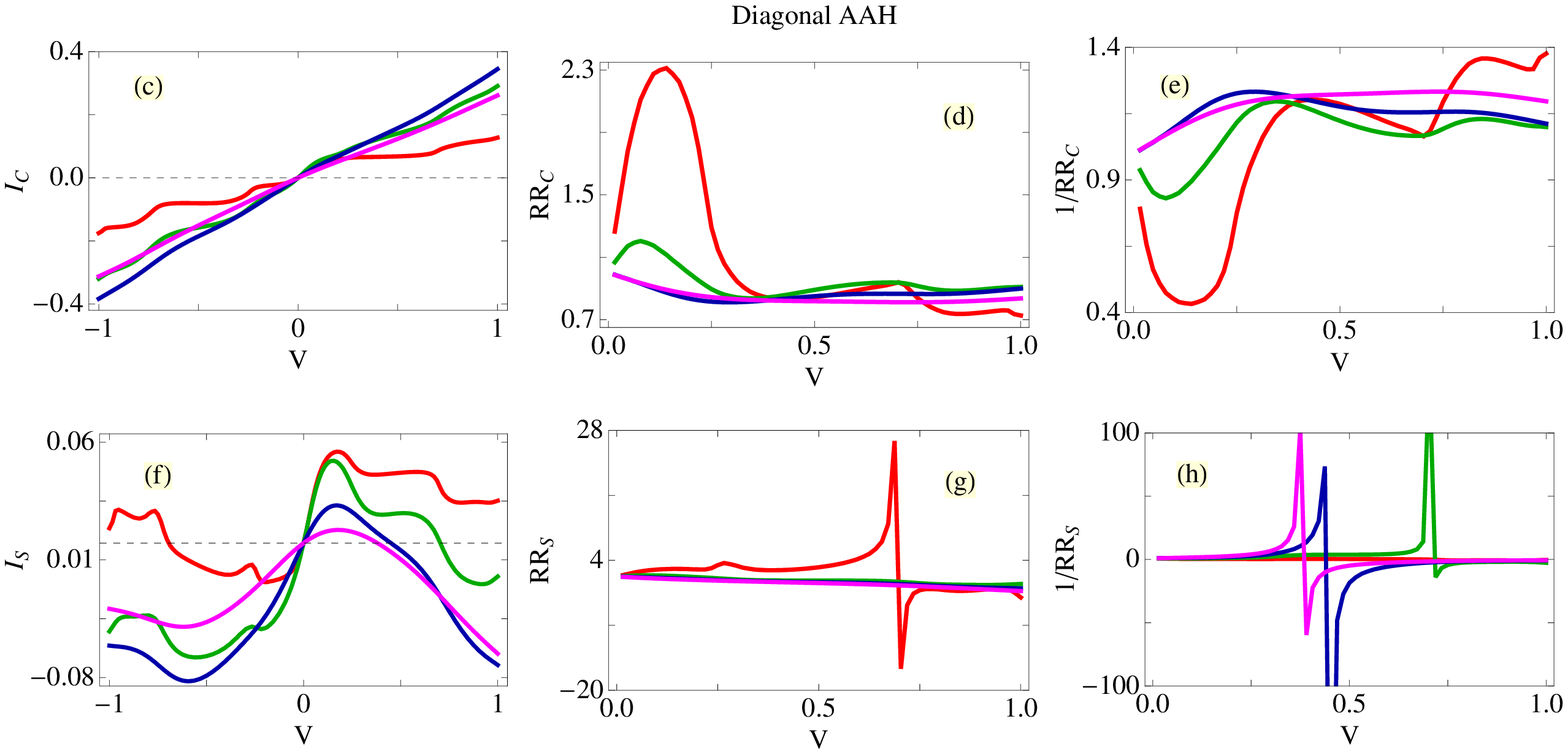}}}\par}
\caption{(Color online). Effect of dephasing (strength is measured by the 
parameter $\gamma$) on different currents and rectification ratios for ordered 
(a-b) and diagonal AAH (c-h) chains. For the AAH chain we choose $\nu=0.5$, 
$\lambda=0$, and $\phi_{\nu}=0$. The other common parameters are: $N=20$ and 
$h=0.5$.}
\label{fig9}
\end{figure*}

\subsection{Generalized AAH chain}

The above analysis shows that the off-diagonal AAH model is quite trivial since
on one hand it does not provide charge current rectification, and on the other 
hand, rectification ratio for spin current cannot be tuned anymore with the phase
$\phi_{\beta}$. But it seems trivial only due to the fact that the diagonal part
is uniform. With the inclusion of incommensurate modulation in site energies, 
interesting phenomena can be expected. To reveal this fact look into the spectra
given in Fig.~\ref{fig8}, where we present the rectification ratios of charge and
spin currents considering a generalized AAH chain, and establish the critical 
roles of two phases $\phi_{\nu}$ and $\phi_{\lambda}$. The one important 
observation is that a high degree of charge current rectification is obtained
and the maximum of it reaches almost close to $15$, which is much higher compared
to the diagonal AAH model (see Figs.~\ref{fig6}). For the case of spin current,
we get quite analogous behavior like what we get in the case of diagonal AAH 
model. At some typical phases $RR_s$ or $1/RR_s$ is too large that we cut the
peaks after a certain value for better presentation. 

The key signature of this
generalized AAH model is that, both charge and spin current rectifications can 
be controlled by tuning the diagonal and off-diagonal phases. As these phases
can be regulated independently with a suitable setup, we can explore their 
combined effects to design an efficient nanoscale rectifier that will be used 
to rectify charge current and spin current as well in a tunable way. 

\subsection{Dephasing effect}

Finally, we focus on the dephasing effect~\cite{dpr1,dpr2,dpr3} which is very 
relevant both in the contexts of practical applications and the fundamental 
points of view. The main essence is to check whether the results studied here 
in absence of dephasing still persist and any other non-trivial features appear 
in presence of dephasing. Many possible sources are there that may destroy phase 
and spin memory of electrons, and among them the most common source is 
electron-phonon (e-ph) interaction. From the measurement of vibrational spectrum 
through inelastic tunneling spectroscopy~\cite{dphexp1,dphexp2} it is possible 
to infer the strength of e-ph coupling. Roughly it is analogous to finding the 
position of the peaks in second-order derivative of $I$-$V$ curve, where the 
voltages associated with these peaks illustrate eigenenergies of the phonons. 
The main challenge is how to incorporate this effect in analyzing electron 
transport. Though several methods are available essentially based on density 
functional theory (DFT) along with non-equilibrium Green's function (NEGF) 
formalism~\cite{dft1,dft2}, but most of them are too heavy and time consuming,
as they require self-consistent solutions. One can circumvent these expensive 
methods and quantitatively explain the basic mechanisms of dephasing on transport
properties by introducing the phenomenological voltage probes into the system.
This is the well known B\"{u}ttiker's scattering approach~\cite{dpmtd1,dpmtd2,
dpmtd3,dpmtd4,dpmtd5,dpmtd6}, where virtual probes are incorporated at each 
sites of the conductor. As these are voltage probes, they do not carry any net 
current and they are responsible to destroy the phase memory of charge carriers. 
One may also use another prescription by considering reduced density matrix 
elements where equation of motions are illustrated in terms of Redfield 
equation~\cite{red1,red2,red3}, but due to enormous simplicity and especially 
the use of minimum physical parameters, here we use B\"{u}ttiker probe method 
to describe the dephasing effects. 

We consider the virtual probes similar to real electrodes and connect them at
different sites of the bridging conductor through the coupling parameter 
$\gamma$. In order to set the condition that these electrodes (virtual) are
not carrying any net current, we need to choose the chemical potentials in such
a way that the voltage drop across each such electrodes is zero. That can be
done by applying a voltage across the real electrodes, viz, $V_S=V_0$ (say) 
and $V_D=0$. Then the effective transmission probability becomes:
$T_{eff}^{\sigma\sigma^{\prime}}=T_{SD}^{\sigma\sigma^{\prime}} +
\sum_p T_{pD}^{\sigma\sigma^{\prime}} V_p/V_0$. 

Now come to the results shown in Fig.~\ref{fig9}, where both the ordered and 
diagonal AAH chains are taken into account. For the ordered case, we get usual 
behavior of charge current, identical magnitudes in two bias polarities, and 
the overall current amplitude gets suppressed with increasing the strength of
the dephasing parameter $\gamma$ (Fig.~\ref{fig9}(a)). On the other hand, a
complete phase reversal takes place in spin current with the inclusion of 
dephasing, and some enhancement is also obtained. But, for this ordered chain 
as current magnitudes are always same in two bias polarities we cannot expect 
any rectification in charge current ($RR_c=1$), and full wave rectification 
is obtained for spin current ($RR_s=-1$). 

More interesting results are obtained for the AAH chain. Though, the degree of
rectification gets reduced for charge current in most of the voltage regions with
the addition of dephasing mechanism, for the case of spin current the scenario
is quite different. From Fig.~\ref{fig9}(g), apparently it seems that $RR_s$
decreases with increasing $\gamma$, but if we take the inverse of $RR_s$ then
we see that the ratio is too high for all the dephasing strengths compared to 
the dephasing-free AAH chain. So there is absolutely a finite probability to 
get much higher rectification even in presence of dephasing, and the underlying
physics for all these phenomena lies in the asymmetric nature of density of 
states for up and down spin bands.

\section{Summary and Outlook}

In summary, we have explored the possibilities of getting high degree of charge 
and spin current rectifications in a one-dimensional systems with quasiperiodic
modulations. The quasiperiodicity is introduced in site energies and/or NNH 
integrals in the form of AAH model. Each site of the system sandwiched between
source and drain electrodes is subjected to a finite magnetic moment which 
separates the up and down spin channels. As the system exhibits gapped energy
energy spectrum in presence of AAH modulations, we get a large degree of 
rectification both in charge and spin currents in presence of an external 
electric field associated with the voltage bias. Unlike charge current which
shows only positive rectification, we get both positive and negative 
rectifications in the case of spin current. We were also able to observe half 
and full wave rectifications, and most importantly, we get these two operations
in a single system. Moreover, we have also discussed thoroughly 
how to tune the 
direction and the degree of rectification by means of both diagonal and 
off-diagonal AAH phases, which are extremely important factors for efficient 
designing a device. Finally, we have investigated the role of dephasing by 
incorporating B\"{u}ttiker probes and found that all the characteristic features 
still persist even in presence of dephasing, which essentially gives us a 
confidence that the results presented here can be tested experimentally.

Before an end, we would like to note that to realize these 
models we may think 
about the quantum Hall system i.e., a 2D lattice in presence of transverse magnetic
field where each lattice site is subjected to a finite magnetic moment. The 
2D quantum Hall system exactly maps with the AAH model~\cite{topeqi}. Suitably 
tuning the magnetic field, we can easily modulate the quasiperiodicity. Here 
one can safely ignore the Zeeman term due
to the interaction of magnetic moment with magnetic field, as band separation
between two spin electrons already takes place due to the spin dependent 
scattering term in Eq.~\ref{eq2}, and at the same time Zeeman term is too weak
compared to this scattering. Some other prescription may be available soon to
examine this model in a suitable laboratory setup. Our results undoubtedly 
provide some important inputs towards rectifications at nanoscale level using
quasiperidic systems due to their unique gapped energy spectra.

\section{Acknowledgments}

MP is thankful to the financial support of University Grants Commission, India 
(F. 2-10/2012(SA-I)) for conducting her research fellowship, and SKM would like
to acknowledge the financial support of DST-SERB, Government of India
(Project File Number: EMR/2017/000504).

\appendix
\section{Wave-guide theory to evaluate transmission and reflection 
probabilities through a spin polarized nano-junction}
\label{transp}

\begin{figure}[ht]
{\centering \resizebox*{8cm}{1.8cm}{\includegraphics{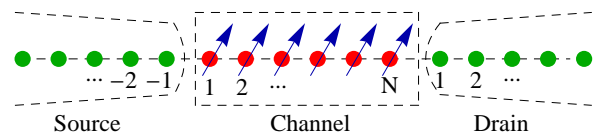}}\par}
\caption{(Color online). Junction setup of the spin polarized device.}
\label{figapen}
\end{figure}

To calculate reflection and transmission probabilities through 
the junction setup given in Fig.~\ref{figapen}, let us begin with the 
wavefunction of the nano-junction 
\begin{equation}
|\psi\rangle =\left[\sum\limits_{n \le -1}^{-\infty}\mbox{\boldmath $A_n$}
\mbox{\boldmath $a_{n,\sigma}^{\dagger}$} + \sum\limits_{n \ge 1}^{\infty}
\mbox{\boldmath $B_n$}\mbox{\boldmath $b_{n,\sigma}^{\dagger}$}
+ \sum\limits_{i=1}^N\mbox{\boldmath $C_i$}\mbox{\boldmath
$c_{i,\sigma}^{\dagger}$}\right]|0\rangle
\label{eqn6}
\end{equation}
where the factors $\mathbf{A_n}$, $\mathbf{B_n}$ and $\mathbf{C_n}$
are expressed as:
\vskip 0.3cm
\noindent
$\mbox{\boldmath $A_{n}$}=\left(\begin{array}{cc}
    A_{n,\uparrow} \\ 
    A_{n,\downarrow}
\end{array}\right)\,$, $\mbox{\boldmath $B_{n}$}=\left(\begin{array}{cc}
    B_{n,\uparrow} \\ 
    B_{n,\downarrow}
\end{array}\right)\,$, and $\mbox{\boldmath $C_{n}$}=\left(\begin{array}{cc}
    C_{n,\uparrow} \\ 
    C_{n,\downarrow}
\end{array}\right)\,$.
\vskip 0.3cm
\noindent
The coefficients $A_{n,\sigma}$, $B_{n,\sigma}$ and $C_{n,\sigma}$ represent
the wave amplitudes of an electron having spin $\sigma$ ($\uparrow$,
$\downarrow$) at $n\,$th site (say) of the source, drain, and $i\,$th 
site (say) of the channel, respectively. $|0\rangle$ is the null state.

Using Eq.~\ref{eqn6}, a set of coupled linear equations are formed from 
the time-independent Schr\"{o}dinger equation $\bf{H}|\psi\rangle$ $=$
$E\bf{I}|\psi\rangle$ ($\mathbf{I}$ being the ($2\times2$) identity matrix),
and they look like,
\begin{eqnarray}
\left(E\mathbf{I}-\mbox{\boldmath $\epsilon_{0,\sigma}$}\right)
\mathbf{A_n}&=&\mbox{\boldmath $t_{0,\sigma}$}\left(\mathbf{A_{n+1}} +
\mathbf{A_{n-1}}\right), n \leq -2, \nonumber \\
\left(E\mathbf{I}-\mbox{\boldmath $\epsilon_{0,\sigma}$}\right)
\mathbf{A_{-1}}&=&\mbox{\boldmath $t_{0,\sigma}$}\mathbf{A_{-2}} +
\mbox{\boldmath $t_{S,\sigma}$}\mathbf{C_1},\nonumber \\
\left(E\mathbf{I}-\mbox{\boldmath $\epsilon_{0,\sigma}$}\right)
\mathbf{B_n} &=&\mbox{\boldmath $t_{0,\sigma}$}\left(\mathbf{B_{n+1}} +
\mathbf{B_{n-1}}\right),n\geq2,\nonumber \\
\left(E\mathbf{I}-\mbox{\boldmath $\epsilon_{0,\sigma}$}\right)\mathbf{B_1}
&=&\mbox{\boldmath $t_{0,\sigma}$}\mathbf{B_2} + 
\mbox{\boldmath $t_{D,\sigma}$}\mathbf{C_N},\nonumber \\
\left(E\mathbf{I}-\mbox{\boldmath $\epsilon_{i,\sigma}$}\right)
\mathbf{C_i} & = &\mbox{\boldmath $t_{i,\sigma}$}\left(\mathbf{C_{i+1}} +
\mathbf{C_{i-1}}\right)
+ \mbox{\boldmath $t_{S,\sigma}$}\delta_{i,1}\mathbf{A_{-1}} \nonumber \\
 & & +~ \mbox{\boldmath $t_{D,\sigma}$}\delta_{i,N}\mathbf{B_1},~ 1\leq
i\leq N
\label{eqn7}
\end{eqnarray}
\noindent
(i) \underline{Incidence of up spin electrons from the source end}
\vskip 0.2cm
Since the source and drain electrodes are perfect, we can write the wave 
amplitudes when up spin electrons are injected from the source end as:
\vskip 0.3cm
\noindent
$\mbox{\boldmath $A_n$}=\left[\begin{array}{cc}
    e^{ik(n+1)a} + r_{\uparrow\uparrow}e^{-ik(n+1)a} \\
    r_{\uparrow\downarrow}e^{-ik(n+1)a} 
\end{array}\right]$
and \\ 
    $\mbox{\boldmath $B_n$}=\left[\begin{array}{cc}
    t_{\uparrow\uparrow}e^{ikna} \\
    t_{\uparrow\downarrow}e^{ikna} 
\end{array}\right]\,$
\vskip 0.3cm
\noindent
where, $k$ is the wave vector associated with the injecting electron energy 
$E$ and $a$ is the lattice spacing. The other coefficients are as follows:
\vskip 0.3cm
\noindent
$t_{\uparrow\uparrow}$ = transmission amplitude of a up spin transmitted 
as up spin, \\
$t_{\uparrow\downarrow}$ = transmission amplitude of a up spin transmitted 
as down spin, \\
$r_{\uparrow\uparrow}$ = reflection amplitude of a up spin reflected 
as up spin, \\
$r_{\uparrow\downarrow}$ = reflection amplitude of a up spin reflected 
as down spin.
\vskip 0.3cm
\noindent
Now plugging $\mathbf{A_n}$ and $\mathbf{B_n}$ into the set of coupled
difference equations given in Eq.~\ref{eqn7} and solving them, we calculate
the spin dependent reflection and transmission amplitudes for each $k$,
associated with the energy $E$. Finally, we evaluate pure spin transmission 
and spin flip transmission probabilities from the expressions
$T_{\uparrow\uparrow}=|t_{\uparrow\uparrow}|^2$ and 
$T_{\uparrow\downarrow}=|t_{\uparrow\downarrow}|^2$, respectively.

\vskip 0.2cm
\noindent
(ii) \underline{\em{Incidence of down spin electrons from the source end}}
\vskip 0.3cm
For the case of down spin incidence the amplitudes $\mathbf{A_n}$ and
$\mathbf{B_n}$ get the forms:
\vskip 0.3cm
\noindent
$\mbox{\boldmath $A_n$}=\left[\begin{array}{cc}
    r_{\downarrow\uparrow}e^{-ik(n+1)a} \\
        e^{ik(n+1)a} + r_{\downarrow\downarrow}e^{-ik(n+1)a}
    \end{array}\right]$
and \\
$\mbox{\boldmath $B_n$}=\left[\begin{array}{cc}
    t_{\downarrow\uparrow}e^{ikna} \\
    t_{\downarrow\downarrow}e^{ikna} 
    \end{array}\right]\,$
\vskip 0.3cm
\noindent
where the meaning of different quantities are:
\vskip 0.3cm
\noindent
$t_{\downarrow\uparrow}$ = transmission amplitude for down spin
transmitted as up spin, \\
$t_{\downarrow\downarrow}$ = transmission amplitude for down spin
transmitted as down spin, \\
$r_{\downarrow\uparrow}$ = reflection amplitude for down spin
reflected as up spin, \\
$r_{\downarrow\downarrow}$ = reflection amplitude for down spin reflected 
as down spin.

\vskip 0.3cm
\noindent
In the same fashion as described above for the case of up spin electrons, 
we determine the reflection and transmission amplitudes for the incidence
of down spin electrons using Eq.~\ref{eqn7}, and eventually, find the 
transmission probabilities as 
$T_{\downarrow\downarrow}=|t_{\downarrow\downarrow}|^2$ and
$T_{\downarrow\uparrow}=|t_{\downarrow\uparrow}|^2$.

In a similar way, one can calculate the spin dependent reflection
probabilities. The method of wave-guide theory described here is the most 
general one, and can easily be utilized to investigate spin dependent 
transmission and reflection probabilities through any spin polarized 
nano-junction.

\end{document}